\documentclass[english,reqno,dvips,12pt]{article}
\usepackage[T2A]{fontenc}
\usepackage[cp1251]{inputenc}
\usepackage[english]{babel}
\usepackage{epsfig,graphicx,amsmath,amsfonts,verbatim,amssymb,amsthm}

\theoremstyle{plain}

\theoremstyle{remark}

\begin{document}

\title{\textbf{Thermodynamic Concept of  Neutron Separation  Energy}}

\author{\textbf{V.P. Maslov}
\thanks{ National Research University Higher School of Economics, Moscow, 123458, Russia;
Moscow State  University,  Physics  Department,  Moscow, 119234, Russia;
 }
}
\date{ }

\maketitle


\begin{abstract}

In the paper, we propose a new approach to the mathematical description of the separation of a neutron from the atom's nucleus on the
basis of the formalisms of tropical mathematics and nonstandard analysis. In studying the behavior of individual nucleons of the
atom's nucleus, instead of the ordinary approach involving the Einstein formula relating mass and energy, we use the thermodynamical
approach to the disintegration of the nucleus. This approach allows to obtain a previously unknown general expression for the energy
needed to separate a neutron from its atom's nucleus, provided we know the de Broglie wavelength and the volume of the nucleus.

\end{abstract}

\section*{Introduction}

At the present time, problems which arose when quantum mechanics arose have returned to the frontline of current research. For
example, the Einstein--Podolsky--Rosen paradox ~\cite{Bell}, the de Broglie wave-particle dualism~\cite{de_Broglie}, and other
problems. Modern mathematics has moved forward since the appearance of these problems, and has led to the solution of part of them.
In particular, this can be said of nonstandard analysis and the dequantization procedure related to tropical
mathematics~\cite{Litvinov_Mas_dequan}. In the present paper, we solve the problem of calculating the specific energy of the
disintegration of a Bose particle.

The creation of contemporary wave mechanics begins with de Broglie's note ``Ondes et Quanta'', presented in 1923. In it, de Broglie
studied the motion of  electrons along a closed orbit and showed that the requirement that the phases agree leads to the
Bohr--Sommerfeld quantization condition, i.e., to the quantization of the angular momentum (see~\cite{Tyurin}--\cite{Esina_Shafar},
as well as~\cite{Yoshioka}--\cite{Czyz} and ~\cite{Mas_cycle_Baril}--\cite{Mas_cocycle}). Developing his ideas about waves, as
related to particles, in 1927 de Broglie presented his {\it theory of double solution} \cite{de_Broglie}, which led to the
wave-particle dualism, still topical today.

De Broglie came to the conclusion that the presence of a continuous wave is related to the appearance of an extra term in the
expression for the Lagrangian particle; this term can be regarded as a small addendum to the potential energy (compare
formula~\eqref{N01-0} below).

We shall consider the case when the number of molecules in a gas is small. We will look at the neutrons and protons (nucleons) that
constitute the atomic nucleus of molecules from the thermodynamical point of view. In particular, we shall make use of the de Broglie
wavelength, which determines the size of the wave packet constituting the given quantum particle.

The Hartrey--Fock equation describing to the weak interaction near the intersection of the wave packets allows to write out the
self-consistent relation for the value of the potential that suffices to keep the nucleon in the nucleus and impedes the
disintegration of the latter. Our approach consists in applying thermodynamical methods, related to the de Broglie wavelength,
together with mathematical methods in number theory and in nonstandard analysis in order to compute the energy needed to separate the
neutron from the atom's nucleus.

According to wave-particle duality, the corpuscular or wave-like character of a particle may be determined by a quantitative
parameter -- the de Broglie wavelength. If the de Broglie wavelength is relatively large, then the particle is a wave packet, i.e.,
it is quantum. In particular, such quantum particles in nuclear physics are called bosons and fermions.

Under the condition that we know the de Broglie wavelength and the volume of the nucleus, it is possible to determine the energy
required for the neutron  to separate from the nucleus and the latter is transformed from a Bose particle to a Fermi particle or vice
versa. This energy is usually calculated by means of the defect of mass, using the formula relating mass to energy discovered by
Einstein. We will do this differently, namely, by using the de Broglie wavelength, we will determine whether the particle is quantum,
and if so, we will find the level of energy at which the neutron is torn away from the atomic nucleus, thereby changing its spin. If
the number of nucleons is even, then the atomic nucleus is a Bose particle. When one nucleon is torn away from such a nucleus, it
becomes Fermi particle with nonzero spin.

We introduce several notion which we use later.

The nucleus spin $J$ is the angular momentum of the nucleus in the state of rest,
i.e., when its orbital (outer) angular momentum is zero. The spin of a nucleus is the vector sum
of spin $\vec s$ and orbital $\vec L$ momenta of the particles contained in the nucleus,
protons and neutrons.

The total momentum of a separate nucleon is half-integer. Therefore, for a nucleus with even number of nucleons $A$,
the nucleus spin is a certain integer, and for a nucleus with odd $A$, it is a half-integer number.
The experimental measurements confirm this.

A nucleus is a boson if it has an integer spin, and it is a fermion if it has a half-integer spin.

Thus, all nuclei with even number of nucleons are bosons. They can be divided into two groups:
even-even and odd-odd. In even-even nuclei, the number of protons is even, as well as the number of neutrons.
In odd-odd nuclei, the number of nucleons of both types is odd.

Examples of odd-odd bosons are: boron-10 ($\mathcal{Z}=5$, $J=3$),  where $\mathcal{Z}$ is a charge number (the number of protons), lithium-6 ($\mathcal{Z}=3$, $J=1$), nitrogen-14 ($\mathcal{Z}=7$, $J=1$).
The even-even bosons are, for example, helium-4 ($\mathcal{Z}=2$, $J=0$), neon-20 ($\mathcal{Z}=10$, $J=0$),
sulphur-18 ($\mathcal{Z}=16$, $J=0$).

All odd nuclei are fermions.
Examples of fermions are: fluorine-19 ($\mathcal{Z}=9$, $J=1/2$),
aluminum-27 ($\mathcal{Z}=13$, $J=5/2$), bromine-81 ($\mathcal{Z}=35$, $J=3/2$).

The even-even nuclei are most stable with respect to the separation of a neutron.
In such nuclei, the energy required to split (separate) a neutron is anomalously high.
When passing to the neighboring nucleus that is odd in neutrons,
it decreases by 10--15\,MeV. This means that to turn a boson into a fermion
is more difficult than conversely.

The spin $J$ of even-even nuclei is always equal to zero.
This is an experimentally confirmed fact, which is explained
by the presence of the so-called coupling forces. Namely, in atomic nuclei,
there arises an additional bond between two nucleons of the same type (two protons or two neutrons)
which occupy the same energy level. Such a pair of nucleons has the maximally possible set of coinciding
quantum numbers and, respectively, the wave functions of nucleons in this pair
are characterized by the greatest overlapping.
The resultant total momentum of such a state of two nucleons is zero.
In even-even nuclei, all nucleons in the ground state are coupled,
which implies that the total momentum of the nucleus is zero.

One of the earliest models of the atomic nucleus was proposed by Niels Bohr in 1936
in the framework of the theory of compound nucleus~\cite{Bohr}.
Later, Carl Weizs\"acker first obtained a semi-empirical formula for the binding energy of the atomic nucleus~$E_c$:
\begin{equation}\label{Vaiczekker}
E_c=\alpha A - \beta A^{2/3} - \gamma\frac{\mathcal{Z}^2}{A^{1/3}} - \varepsilon \frac{(A/2-\mathcal{Z})^2}{A}+\delta,
\end{equation}
where
$$
\delta = \left\{
\begin{aligned}
&+\chi A^{-3/4}  & \, \text{for even-even nuclei},\\
\,\,\, &0        & \, \text{for nuclei with odd $A$},\\
&-\chi A^{-3/4}  & \, \text{for odd-odd nuclei},\\
\end{aligned}
\right.
$$
$A$ is the mass number (total number of nucleons) in the nucleus,
$\mathcal{Z}$ is the charge number (number of protons) in the nucleus, and
$\alpha$, $\beta$, $\gamma$, $\varepsilon$, and $\chi$
are parameters obtained by statistical treatment of experimental data.
This formula provides  sufficiently exact values of the binding energies and masses for many nuclei,
which permits using it to analyze different properties of the atomic nucleus.
In what follows, we obtain new important relations between the temperature and the chemical potential
in the process of nucleon separation from the atomic nucleus.
For this, we use the parastatistic relations modified by a nonstandard analysis and
the mathematical notion of infinitesimals.

\bigskip

\textbf{a. {Bose statistics, Fermi statistics in Hougen--Watson diagrams and in Gentile statistics }}

The behavior of Bose particles and Fermi particles is described by the Bose--Einstein and Fermi--Dirac distributions respectively.

Let $a =e^{\mu/T}$ be the activity ($\mu$ being the chemical potential, $T$ being the temperature).  Let $D$ be the number of degrees
of freedom  (the dimension) and  let $s=D/2$.  The total energy of all $N$ particles  is denoted by $E$.

The Bose--Einstein distribution in terms of the polylogarithm has the form
 \begin{equation}\label{B-E}
    \operatorname{Li}_s(a) = \frac{1}{\Gamma(s)} \int_0^\infty \frac{t^{s-1}}{e^t/a-1}\, dt.
      \end{equation}
The Fermi--Dirac distribution is written as
 \begin{equation}\label{F-D}
   - \operatorname{Li}_s(-a) = \frac{1}{\Gamma(s)} \int_0^\infty \frac{t^{s-1}}{e^t/a+1}\, dt.
      \end{equation}
In these formulas,  $\operatorname{Li}_{(\cdot)}(\cdot)$ is the polylogarithm function,  $\Gamma$ is the Euler gamma-function,
and $t$ is time.

Usually, in physics, the Bose--Einstein and the Fermi--Dirac distributions  are defined with the help of Gentile
statistics~\cite{Gentile}. Gentile statistics contains Bose--Einstein and the Fermi--Dirac statistics as particular cases. Gentile
statistics contains an additional constant $K$ that indicates the maximal number of particles located at a fixed level of energy. In
particular, for $K=1$, the distribution of Gentile statistics coincides with the distribution of the Fermi--Dirac statistics, so that
the corresponding formulas coincide in form with~\eqref{F-D}. In Gentile statistics, we always have $K\geq1$.

For an ideal gas of dimension $D$ obeying the Gentile statistics, i.e., in the case when there are no more than $K$ particles ($K$
being  an integer) at each energy level, the total number of particles $N$ is known:
\begin{equation}\label{Gent}
N=\frac{V}{\Lambda^{2s}}\bigg(\operatorname{Li}_{s}(a) -\frac{1}{(K+1)^{s-1}}\operatorname{Li}_{s}(a^{K+1})\bigg) ,
\end{equation}
where  $V$  is the  volume, $\Lambda$ is the de Broglie wavelength.

In our approach, unlike the standard Gentile statistics, we  set $K=0$. Note that in thermodynamics $N$ means the number of
particles. In the present paper, we do not consider molecules, we only consider nuclei, i.e., we are in nuclear physics. In that
sense, we can say that in our model the number of particles  $N$ is equal to zero.

The consideration of the $\Omega$-potential ({compare}~\cite{Smoczyk}--\cite{Vaisman}) corresponding to Gentile statistics allows us
to describe in more detail the passage from particles of the atomic nucleus of a Bose gas to those of a Fermi gas. In this situation,
by analogy with the $\Omega$-potential considered by Landau and Lifshits~\cite{Landau_StPhys}, this allows to calculate the total
energy of this passage.

We consider quantum particles, each of which is a wave packet~\cite{Littlejohn}. These wave packets are characterized by their de
Broglie wavelength $\Lambda$.

One can see that when the activity changes sign, the distributions~\eqref{B-E} and~\eqref{F-D} also change sign. This corresponds to
the passage from negative pressures to positive ones. This picture naturally arises in the Van-der-Waals
formulas~\cite{MTN_97-6}--\cite{RJMP_22-3}. Thus, the Bose particles and the Fermi particles are positioned in different parts of the
-P,Z-diagram of Hougen and Watson (in it, $P$ is the pressure, $Z=PV/NT$ is the compressibility factor, $V$, the volume, $N$, the
number of particles, $T$, the temperature): the Bose particles are in the positive domain, while the Fermi particles are in the
negative one.

We will use this technique to compute the specific energy required for the passage of particles of an ideal Bose gas to particles of
a Fermi gas. We can consider that the Fermi gas is obtained when the activity changes sign. This process is described by the passage
from formula~\eqref{B-E} to formula~\eqref{F-D}.

\bigskip

\textbf{b. {Gentile statistics}}

Usually, in physics, the Bose--Einstein and the Fermi--Dirac decompositions are defined with the help of Gentile
statistics~\cite{Gentile}. Gentile statistics contains Bose--Einstein and the Fermi--Dirac statistics as particular cases. Gentile
statistics contains an additional constant $K$ that indicates the maximal number of particles located at a fixed level of energy. In
particular, for $K=1$, the distribution of Gentile statistics coincides with the distribution of the Fermi--Dirac statistics, so that
the corresponding formulas coincide in form with~\eqref{F-D}. In Gentile statistics, we always have $K\geq1$.

The consideration of the $\Omega$-potential ({\it compare}~\cite{Smoczyk}--\cite{Vaisman}) corresponding to Gentile
statistics~\cite{Gentile} allows us to describe in more detail the passage from particles of the atomic nucleus of a Bose gas to
those of a Fermi gas. In this situation, by analogy with the $\Omega$-potential considered by Landau and
Lifshits~\cite{Landau_StPhys}, this allows to calculate the total energy of this passage.

We will use use this technique to compute the specific energy required for the passage of particles of an ideal Bose gas to particles
of a Fermi gas. We can consider that the Fermi gas is obtained when the activity changes sign. This process is described by the
passage from formula ~\eqref{B-E} to formula ~\eqref{F-D}.

\bigskip
\textbf{c. {Notation}}

Let us introduce the notation that will allow us to determine the energy in dimensionless form.

Let $\mathbf{v}=\Lambda^{2s}$. This quantity has the dimension of volume in
$2s-$dimensional space. Let
$\mathbf{e}=\frac{2\pi\hbar^2}{m}{V}^{-\frac{1}{s}}$. This quantity has the dimension of energy.

Now let us introduce the dimensionless quantities
$\mathcal{E}={E}/{\mathbf{e}}$ for the energy and
$\mathcal{V}={V}/{\mathbf{v}}$ for the volume. Note that that the quantity
 $\mathcal{V}^{1/D}$ is the ratio of the characteristic linear size of the system
${V}^{1/D}$ to the de Broglie wavelength $\Lambda$.

Usually one denotes by $N_i$ the number of particles located at the $i$-th level of energy. It is customary to consider that in the
case of a Fermi gas there can be no more than one particle at a fixed level, while for a Bose gas the number of particles $N_i$ at a
given level can be as large as we wish. We shall consider the Gentile statistics ~\cite{Gentile}, in which at each energy level there
is no more than $K$ particles. In other words, the number of particles at any level of energy cannot be greater than $K$.

The maximal number of particles at a given energy level occurs when the activity $a$ is maximal, i.e., at the point  $a=1$. Since
$\sum_{i=1}^M N_i = N$, for any Bose system we obviously have $N_i\leq N$. Therefore, for a Bose system, we have $K\leq N$. In
Gentile statistics, $K$ takes integer values.

\section*{Results}

\subsection*{Passage from the Bose-type region to the Fermi-type region. Computing  of $a_0$}

We denote  by $a_0$  the maximal value of the activity $a$ for Bose particles as $N\to 0$.  The quantity $a_0$ indicates the maximal value of the activity at which the decomposition of bosons into fermion occurs.

Let us  obtain an expression for $a_0$.

We will assume that $K=N$ in an infinitely small neighborhood of $[N]$, where
$[N]$ is the integer part of the number $N$.

The set of all points infinitely close to the number $[N]$ is called the Leibnitz differential ~\cite{Shepin-2} in nonstandard
analysis as developed by Robinson (see~\cite{Nestandart-1}--\cite{Nestandart-2}); the Leibnitz differential can be understood as the
length of an elementary infinitely small segment (the monad). A differential is an arbitrary infinitely small increment of the
variable.

In the  positive domain  of the -P,Z-diagram (Bose particles case)  monads, i.e. fractional numbers, vary in the  interval between~0
and~1. In the negative domain (Fermi particles case), they lie in the interval between~$-1$ and~0. Hence, on passing  from
Bose particles to Fermi particles monads take all values from~0 to~$|1|$.

Denote by $x^p$ the difference $N-[N]$, i.e., $N-[N]=x^p<0$ ($x>0$ corresponds to bosons in the -P,Z-diagram). We note that  $p$ is
an arbitrary  number,  including  monads, hyperreals or nonstandard reals, fractal numbers, etc. Hence the  equations corresponding
to the energy under consideration  do not depend on~$p$.

This allows one to draw an analogy between integers in the standard number theory
and their generalization in the abstract analytic number theory (see~\cite{Postnikov}).
This problem was studied in detail in~\cite{Mas_Naz_100-3}--\cite{Mas_Dobr_Naz}.
This corresponds to the generality of computed ``specific''  separation energy of neutrons
from the atomic nucleus.
This approach allows us to assume that if one neutron separates from a boson nucleus, then its spin was equal to zero,
and if one neutron separates from a fermion nucleus, then its spin becomes equal to zero after separation.
In this sense, the methods of tropical geometry can be transferred to the  mathematical analysis
(see~\cite{Litvinov_Mas_dequan}).

We shall search for the expansion in powers of
$x$ up to $O(x^{2p})$, which will imply that $N\sim [N]$.

The self consistent relation for $x$  in a neighborhood of $[N]$ is of the form:
\begin{equation}\label{Np}
[N]+x^p=\frac{{V}/\Lambda^{2s}}{\Gamma(s)}
\int_0^\infty\bigg(\frac{1}{e^{\xi}/a-1}-\frac{[N]+x^p+1}{e^{([N]+x^p+1)\xi}/a-1}\bigg)  \xi^{s-1} \, d\xi,
\end{equation}

The following thermodynamical formula for the energy is known:
\begin{equation}\label{Mp}
\mathcal{E}=s\frac{(V/\Lambda^{2s})^{\frac{s+1}{s}}}{\Gamma(s+1)}
\int_0^\infty\bigg(\frac{1}{e^{\xi}/a-1}-\frac{[N]+x^p+1}{e^{([N]+x^p+1)\xi}/a-1}\bigg)  \xi^{s} \, d\xi,
\end{equation}

Let us perform the same manipulations for $x>0$. Then the term at the first power of $x$ will be negative. It corresponds to a Fermi
system.

Therefore, in our approach, unlike the standard Gentile statistics, we also set $K=0$ and only consider the case $[N]=0$. To the
numbers $N=K=0$, we apply nonstandard analysis as well as the technique of Gentile statistics~\cite{Gentile}.

Using the technique of nonstandard analysis, we add to the integer $K$ the monad $x$.
Then the expression~\eqref{Gent} is not equal to zero.

Let us expand the right-hand side of equation~\eqref{Np} in powers of small values of $x$, neglecting terms of degree $3p$ or higher:
\begin{equation}     \label{N01-0}
\begin{split}
&x^p=\frac{V}{\Lambda^{2s}} x^p \bigg((s-1) \operatorname{Li}_{s}(a)-\log (a)\operatorname{Li}_{s-1}(a)\bigg)\\
&+\frac{V}{\Lambda^{2s}} \frac{1}{2} x^{2p} \bigg(\log ^2(a) (-\operatorname{Li}_{s-2}(a))-(s-1)
\big(s \operatorname{Li}_{s}(a)-2 \log (a)\operatorname{Li}_{s-1 }(a)\big)\bigg)+....
\end{split}
\end{equation}

Dividing both parts of~\eqref{N01-0} by $x^p$ we obtain
\begin{equation}     \label{N01-1}
\begin{split}
&1=\frac{V}{\Lambda^{2s}}  \bigg((s-1) \operatorname{Li}_{s}(a)-\log (a)\operatorname{Li}_{s-1}(a)\bigg)\\
&
+\frac{V}{\Lambda^{2s}} \frac{1}{2} x^{p} \bigg(\log ^2(a) (-\operatorname{Li}_{s-2}(a))-(s-1)
\big(s \operatorname{Li}_{s}(a)-2 \log (a)\operatorname{Li}_{s-1 }(a)\big)\bigg)+....
\end{split}
\end{equation}

Thus,  we obtain an expression for $a_0$, the value of $a$ for which we have $N=0$ in the Bose--Einstein distribution:
\begin{equation}
\label{N=0}
(s-1) \operatorname{Li}_{s}(a_0)-  \log (a_0)\operatorname{Li}_{s-1}(a_0)-\big(\frac{V}{\Lambda^{2s}}\big)^{-1}=0.
\end{equation}

Similarly for the Fermi--Dirac distribution, we find:
\begin{equation}
\label{N=0F}
(s-1)\left(- \operatorname{Li}_{s}(-a_0)\right)
-\log (-a_0)\left(-\operatorname{Li}_{s-1}(-a_0)\right)+\big(\frac{V}{\Lambda^{2s}}\big)^{-1}=0, \qquad a_0 <0.
\end{equation}

The value of $\operatorname{Li}_{s}(a)$, where $a=e^{\mu/T}$, corresponds to the total energy of passage, in particular in the
three-dimensional case ($s=3/2$).

Equation~\eqref{N=0} for sufficiently large values of
$\frac{V}{\Lambda^{2s}} $ has a unique solution $a_0\le1$, which depends on  $\frac{V}{\Lambda^{2s}}$ and $s$.

The value of the activity $a$ for a known temperature $T$ gives the corresponding value of the chemical potential~$\mu$:
\begin{equation}
\label{mu0}
\mu=T \log(a)\le0.
\end{equation}

In particular, for $a=a_0$, the higher the temperature $T$, the smaller is $a_0$ and the larger becomes the corresponding value of
$|\mu_0|$. Thus, as the temperature grows, the point of passage $\mu_0$ approaches the point $\mu=-\infty$, at which the pressure $P$
changes sign.

Let $a_0=1$ and assume that we know the mass $m$ of one nucleon and the volume $V$ of the nucleus. Then equation~\eqref{N=0} with the
value $\Lambda=\sqrt{\frac{2\pi\hbar^2}{m T}}$ of the de Broglie wavelength taken into account can be regarded as an equation for the
unknown $T$.

Let us call {\it critical} the temperature that arises for $a_0=1$, i.e., as $\mu_0 \to 0$. Denote this temperature by~$T_s$.

From~\eqref{N=0} for $s>1$ we obtain
\begin{equation}
\label{Ts}
T_s=\frac{2 \pi \hbar^2}{m (V (s-1) \zeta(s))^{1/s}},
\end{equation}
where $\zeta(\cdot)$ is the Riemann zeta function.

Since the temperature $T_s$ is the lowest one in the whole range of variation of $\mu_0$ (which is the ray $(-\infty,0]$), we will
call the ratio $T/T_s$ the {\it regularized} temperature and denote it by  $T_{\text{reg}}$. The change of temperature
 can be measured by $T_{\text{reg}}$.

Expanding the energy \eqref{Mp} in powers of $x$ to the first degree inclusive,  we obtain:
\begin{equation}
\label{E1}
\mathcal{E}_{sp}(dx)^p=
2s \big(\frac{V}{\Lambda^{2s}}\big)^{\frac{1}{s}+1}
\bigg(s \operatorname{Li}_{s+1}(a_0)-\log (a_0)\operatorname{Li}_{s}(a_0)\bigg) (dx)^p.
\end{equation}
The value of  $a_0$ can be computed from~\eqref{N=0} and \eqref{N=0F}.

Thus we have calculated the specific  energy needed to separate a neutron from an atomic nucleus when one neutron leaves the nucleus,
provided the volume of the nucleus and the de Broglie wavelength are known.

Let us consider the case of parastatistics with infinitely small $K$ and $N$ equal to each other.

In the case of parastatistics, we have the following relations,
in which the first term in parentheses gives the distribution for Bose particles,
and the second term, the parastatistical correction:
\begin{equation}\label{ep}
E=\frac{V}{\lambda^{D}}T({\gamma+1}) (\operatorname{Li}_{2+\gamma}(a)-\frac{1}{(K+1)^{\gamma+1}}\operatorname{Li}_{2+\gamma}(a^{K+1})),
\end{equation}
\begin{equation}\label{np}
N= \frac{V}{\lambda^{D}} (\operatorname{Li}_{1+\gamma}(a)-\frac{1}{(K+1)^{\gamma}}\operatorname{Li}_{1+\gamma}(a^{K+1})),
\end{equation}
where $\operatorname{Li}_{(\cdot)}(\cdot)$  is the polylogarithm function,
$a =e^{\mu/T}$ is the activity ($\mu$ being the chemical potential, $T$ being the temperature),
$\gamma=D/2-1$, $D$  is the number of degrees of freedom (the dimension),
$\lambda=\sqrt{\frac{2\pi\hbar^2}{m T}}$ is the de Broglie wavelength,
where $\hbar$ is the Planck constant, $m$ the mass of one particle.

If in  the intermediate region between Bose particles and Fermi particles there is a certain number of nucleons, then physicists
refer to it as a nuclear halo. The region corresponding to the difference of pressure $P=0$ and to an infinitely small sequence
$\{P_K\}\to 0$ constitutes the nuclear halo. The passage from the Bose-type region to the Fermi-type region occurs through the
nuclear halo,  which  contains the value of the pressure $P=0$.  Above we denoted  by $a_0$  the maximal value of the activity $a$
for Bose particles as $N\to 0$. The quantity $a_0$ indicates the maximal value of the activity at which the decomposition of bosons
into fermion occurs.

For an ideal gas of dimension $D=3$, relations  \eqref{ep}, \eqref{np} become
\begin{equation}\label{Gent-1}
N=\frac{V}{\lambda^{3}}(\operatorname{Li}_{3/2}(a) -\frac{1}{(K+1)^{1/2}}\operatorname{Li}_{3/2}(a^{K+1})),
\end{equation}
\begin{equation}\label{Mp-1}
{E}= \frac{3}{2} \frac{{V}}{\lambda^{3}}T(\operatorname{Li}_{5/2}(a)-\frac{1}{(K+1)^{3/2}}\operatorname{Li}_{5/2}(a^{K+1})).
\end{equation}

The expansion of the summand
$\frac{1}{(K+1)^{1/2}}\operatorname{Li}_{3/2}(a^{K+1})$ from formula \eqref{Gent-1} in small values of $K$ has the form:
\begin{equation}     \label{raz3}
\begin{split}
&\frac{1}{(K+1)^{1/2}}\operatorname{Li}_{3/2}(a^{K+1})
=\operatorname{Li}_{3/2}(a)-[K ( \operatorname{Li}_{3/2}(a)/2-\log (a)\operatorname{Li}_{1/2}(a))+O(K^2).\\
\end{split}
\end{equation}
Let $B=V/\lambda^3 > 0$. Then equation~\eqref{Gent-1} for small $K$ acquires the form:
\begin{equation}     \label{N01-0-1}
\begin{split}
&N=B K (\frac{1}{2} \operatorname{Li}_{3/2}(a)-\log (a)\operatorname{Li}_{1/2}(a))+O(K^2).\\
\end{split}
\end{equation}

Dividing both sides of ~\eqref{N01-0-1} by $N$ and taking the limit as $K\to0$, yields an expression for $a_0$, i.e., the value of
$a$ for which $K=N=0$:
\begin{equation}
\label{N=0-2}
\frac{1}{2} \operatorname{Li}_{3/2}(a_0)-  \log (a_0)\operatorname{Li}_{1/2}(a_0)-B^{-1}=0.
\end{equation}

Equation~\eqref{N=0-2} in the case of an arbitrary coefficient $\gamma=D/2-1$ instead of $1/2$,
after similar arguments, acquires the form:
\begin{equation}
\label{N=0-gen-2}
\gamma \operatorname{Li}_{\gamma+1}(a_0)-  \log (a_0)\operatorname{Li}_{\gamma}(a_0)-\frac{\lambda^{2(\gamma+1)}}{V}=0.
\end{equation}

Equation~\eqref{N=0-gen-2} has a unique solution $a_0>0$ that depends on $B$ and $\gamma$.

In the case $K=N$, equation \eqref{Gent-1} acquires the form
\begin{equation}\label{GentKN}
    N=B(\operatorname{Li}_{3/2}(a) -\frac{1}{(N+1)^{1/2}}\operatorname{Li}_{3/2}(a^{N+1})).
\end{equation}

This equation obviously has the solution $N\equiv 0$ for any $a\geq 0$. However, for $a>a_0$, it has one more nonnegative solution
$N(a)$. This can be verified by constructing the graphs of the right-hand and left-hand sides of \eqref{GentKN} as a function of $a$
for an arbitrary fixed $N>0$. The right-hand side of the equation is  zero for $a=0$ and monotonically grows for $a\in(0,\infty)$,
while the left hand side is a constant that does not depend on $a$.

Substituting the obtained relation $N(a)$ in formula \eqref{Mp-1}, we can find the dependence $E(a)$, and with it the pressure
$P(a)$, by using the relation $E=(\gamma+1)PV$.

In our considerations $K$ is an infinitely small number. Thus we are not dealing with the Fermi statistics or the Bose statistics,
but with a parastatistics of a new type, which can be called a Bose-like statistics.

Let us substitute the obtained relation into the graph of the compressibility factor $Z=PV/(NT)$ as a function of $P$ (this graph is
known as the Hougen--Watson diagram).

For Fermi statistics in the case $D=3$, we have the relations
\begin{equation}\label{Nf}
    N=-\frac{{V}}{\lambda^{3}}\operatorname{Li}_{3/2}(-a),
\end{equation}
\begin{equation}\label{Ef}
    E= -\frac{3}{2} \frac{{V}}{\lambda^{3}}T \operatorname{Li}_{5/2}(-a).
\end{equation}

Let us call the curve on the  P-Z diagram, constructed according to formulas \eqref{Nf}--\eqref{Ef} of Fermi statistics, the Fermi
branch. The pressure $P$, as well as the number of particles $N$, on the Fermi branch is positive.

In the Bose-like region, the boson consists of two fermions of the same mass, while in the Fermi-like region the pair of fermions
differ in mass. As the temperature grows, the difference in mass increases and the fermion with the smaller mass disappears. One
fermion remains.

The table~\ref{tabl:t1}  presents values of $a_0$ and $\mu_0= T \log a_0$ for various isotopes. The value of $a_0$ in the case
of separation can be found by means of formula \eqref{N=0-2}, taking into account the expression of the de Broglie wavelength
$\lambda$ in terms of the volume $V$ of the nucleus, its temperature $T$ and its mass $m$. The volume of the nucleus is taken to be
that of a ball of radius  $r_0=A^{1/3} 1.2\times 10^{-15}$ m$^3$. The temperature $T$ of the nucleus expressed in energy units is
taken equal to the energy of separation of a neutron $B_{nExp}$ (obtained from the database CDFE), since it is equal to the
excitation of the nucleus.

\begin{table}
    \caption{Results for isotopes of various  chemical elements}
    \label{tabl:t1}
        \begin{tabular}{llll}
        nucleus                    & $B_{nExp}$,\,MeV & $a_0$                  & $\mu_0$,\,MeV \\
        iron-54                    & 13.378           & 0.0000542755           & -131.391     \\
        neon-20                    & 16.9             & 0.00059614             & -125.483     \\
        chromium-50                & 13.001           & 0.0000704527           & -124.297     \\
        germanium-70               & 11.534           & 0.0000338915           & -118.712     \\
        selenium-76                & 11.155           & 0.0000285583           & -116.721     \\
        chromium-52                & 12.039           & 0.0000718161           & -114.869     \\
        carbon-12                  & 18.722           & 0.00219147             & -114.638     \\
        nickel-60                  & 11.388           & 0.0000529799           & -112.122     \\
        krypton-84                 & 10.521           & 0.0000238875           & -111.966     \\
        zinc-66                    & 11.059           & 0.0000427369           & -111.258     \\
        oxygen-16                  & 15.664           & 0.0012928              & -104.18      \\
        xenon-132                  & 8.937            & 9.06854$\times10^{-6}$ & -103.765     \\
        indium-115                 & 9.037            & 0.000012972            & -101.691     \\
        titanium-50                & 10.94            & 0.0000939594           & -101.443     \\
        copper-65                  & 9.911            & 0.0000534668           & -97.489      \\
        lutetium-175               & 7.667            & 5.40851$\times10^{-6}$ & -92.9818     \\
        lead-208                   & 7.369            & 3.61133$\times10^{-6}$ & -92.3441     \\
        Uup--290                   & 6.772            & 1.68742$\times10^{-6}$ & -90.0155     \\
        molybdenum-100             & 8.291            & 0.0000219127           & -88.9495     \\
        lawrencium-261             & 6.792            & 2.23144$\times10^{-6}$ & -88.3834     \\
        neodymium-150              & 7.381            & 8.75154$\times10^{-6}$ & -85.9612     \\
        plutonium-229              & 6.762            & 3.2006$\times10^{-6}$  & -85.554      \\
        copernicium-285            & 6.472            & 1.90294$\times10^{-6}$ & -85.2499     \\
        californium-250            & 6.625            & 2.60987$\times10^{-6}$ & -85.1724     \\
        argon-40                   & 9.9              & 0.000207475            & -83.9569     \\
        silicon-30                 & 10.61            & 0.00041606             & -82.5955     \\
        mercury-199                & 6.664            & 4.7946$\times10^{-6}$  & -81.6208     \\
        radium-226                 & 6.397            & 3.62971$\times10^{-6}$ & -80.1311     \\
        krypton-85                 & 7.121            & 0.0000440699           & -71.4217     \\
        helium-4                   & 20.578           & 0.0559835              & -59.3202     \\
        nitrogen-14                & 10.554           & 0.00384407             & -58.6931     \\
        copper-75 (1.224 s)        & 6.166            & 0.0000791966           & -58.2291     \\
        molybdenum-101 (14.61 min) & 5.399            & 0.0000433015           & -54.2455     \\
        krypton-94 (212 ms)        & 5.152            & 0.0000570847           & -50.3401     \\
        boron-10                   & 8.437            & 0.0162232              & -34.7715     \\
        argon-50 (170 ns)          & 4.472            & 0.000425087            & -34.7171     \\
        lithium-4                  & 11.452           & 0.193449               & -18.8127     \\
        neon-30 (7.3 ms)           & 3.072            & 0.00360392             & -17.2823     \\
        scandium-60 (3 ms)         & 2.072            & 0.000945668            & -14.4286     \\
        sulfur-45 (68 ms)          & 2.272            & 0.00185767             & -14.2873     \\
        oxygen-23 (82 ms)          & 2.742            & 0.00990298             & -12.6541     \\
        lithium-6                  & 5.665            & 0.205984               & -8.95047     \\
        fluorine-29 (2.5 ms)       & 0.972            & 0.0341069              & -3.28366     \\
        carbon-15 (2.449 s)        & 1.218            & 0.210634               & -1.8972      \\
        boron-14 (12.5 ms)         & 0.971            & 0.468423               & -0.73639     \\
        carbon-17 (193 ms)         & 0.728            & 0.429614               & -0.615063    \\
        lithium-8 (839.9 ms)       & 2.033            & 0.762495               & -0.551266    \\
        berillium-9                & 1.667            & 0.76682                & -0.442595    \\
    \end{tabular}
\end{table}

We have obtained an equation from which we can find the value of $a_0$,
and can determine the temperature $T$ at which this value is attained.

\subsection*{Dependence of the chemical potential on the temperature. Construction of the isotherm }

Let us consider the ratio of the compressibility factor  $Z= PV/(NT)$ to the pressure $P$, i.e.,  $Z/P$.

On any isotherm $T$ is constant. The volume $V$ we consider to be constant, $N=K$ (see~\eqref{GentKN}) is an infinitely small number
and therefore $1/N$ is an infinitely large number. Since the values of $V$ and $T$ are constant on isotherms and $P$ decreases, the
ratio $Z/P$ becomes a constant, depending only on $V$ and $T$, divided by the infinitely small number $N$.

Let $\{P_K\}$ be an infinitely small sequence, coinciding with the infinitely small quantity $K$.

\begin{figure}
    \centering
    \includegraphics[width=0.7\linewidth]{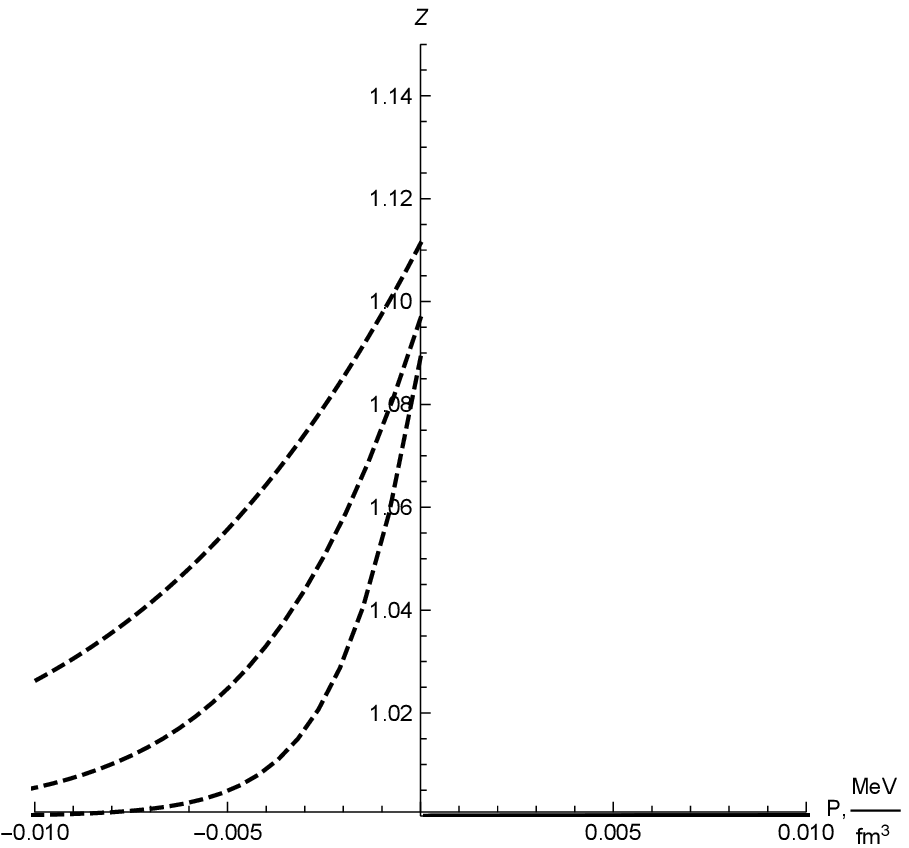}
    \caption{Dependence of the compressibility factor $Z$ on the pressure $P$,
    expressed in the units MeV/fm $^3$ for argon-40, copper-65, and molybdenum-100.
    The continuous line represents the line $Z=1$. The hashed lines show isotherms of the Bose branch,
    constructed according to formulas \eqref{Gent-1}--\eqref{Mp-1}.
    The temperature is equal to the energy needed for the separation of the neutron $B_{nExp}$ (see Table~\ref{tabl:t1}).
    The corresponding value of $a_0$ is given in Table~\ref{tabl:t1}.}
    \label{fig:01}
\end{figure}

Figure~\ref{fig:01} shows the dependence of the compressibility factor on the pressure $P$, expressed in the units MeV/fm$^3$ for
argon-40, copper-65, molybdenum-100. The dashed lines are the isotherms of the Bose branch constructed by means of
formulas~\eqref{Gent-1}--\eqref{Mp-1}. The temperature is equal to the extremal value of the separation energy of the neutron
$B_{nExp}$, indicated in table~\ref{tabl:t1}.

To each value of $a_0$ there corresponds a definite value of the temperature  $T$. In turn, to each value of $T$ there corresponds an
isotherm on the Hougen--Watson diagram. These isotherms lie in the negative quadrant. The temperature characterizing the isotherm
becomes smaller as the point $a_0$ becomes nearer to $Z=1$.

Thus we can say that the Van-der-Waals isotherms are in a sense opposite to the isotherms of nuclear matter shown in
Fig.~\ref{fig:01}.

This shows that the chemical potential $\mu$  at $P=0$ does not become equal to minus infinity and so the  axis  $Z$  at $P=0$ is not
the boundary between  two unrelated structures. Since the value of $|\mu_0|$ is very large but not infinite between the values of the
infinitely small quantities $\{P_K\}$ and the region obeying the Fermi--Dirac distribution, there is a narrow ``halo'' dividing the
Bose region from the Fermi region.

We do not consider the problem of proving that the isotherms in Fig.~\ref{fig:01} exist. We give an approximate solution of the
obtained equations (see similar approaches in  papers~\cite{Bruno},~\cite{Weinstein_Gerbe}), in which all the isotherms corresponding
to different values of $a_0$ are approximately constructed.

\subsection*{Halo nuclei and its fragmentation}

An analysis of Table~\ref{tabl:t1} shows that, as the excitation energy $B_{nExp}$ increases,
the value $a_0$ decreases for almost all nuclei.
This means that the ``distance'' between the Bose- and Fermi-like domains decreases, i.e.,
the halo decreases. In the limit case where the nucleus temperature is infinite,
$a_0=0$ and the halo disappears. Conversely, in the region of small temperatures,
the value $a_0$ is close to the unity, which corresponds to the chemical potential equal to zero.
The last situation corresponds to the case where the halo width is huge
and the Bose- and Fermi-like domains intermix.

\begin{figure}[h!]
    \includegraphics[draft=false]{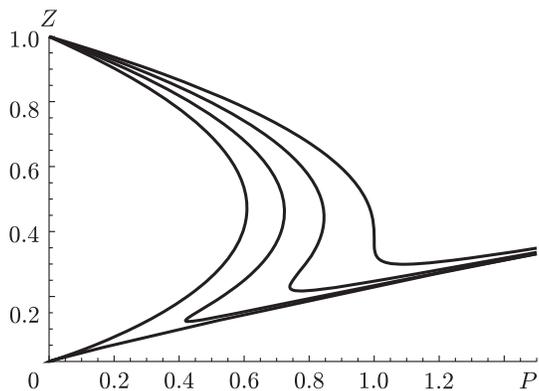}
    \caption{Isotherms $T=0.84375$, $T=0.9$, $T=0.95$, $T=1$  for $\rho>\rho_c$.}
    \label{fig:02}
\end{figure}

In thermodynamics, when the terrestrial attraction is taken into account, the picture changes sharply.
The more liquid falls on the vessel bottom, the heavier is the liquid layer
and the greater is its pressure on the vessel walls.
The higher the water column in the vessel, the greater is the pressure on its walls.
Therefore, the Bose--Einstein formula of the pressure distribution cannot be applied in our case.
It is necessary to take the weight and weak compressibility of the liquid into account.
These elementary considerations permit explaining the isotherms  on the P-Z diagram shown in Fig.~\ref{fig:02}.

\begin{figure}
    \centering
    \includegraphics[width=\linewidth]{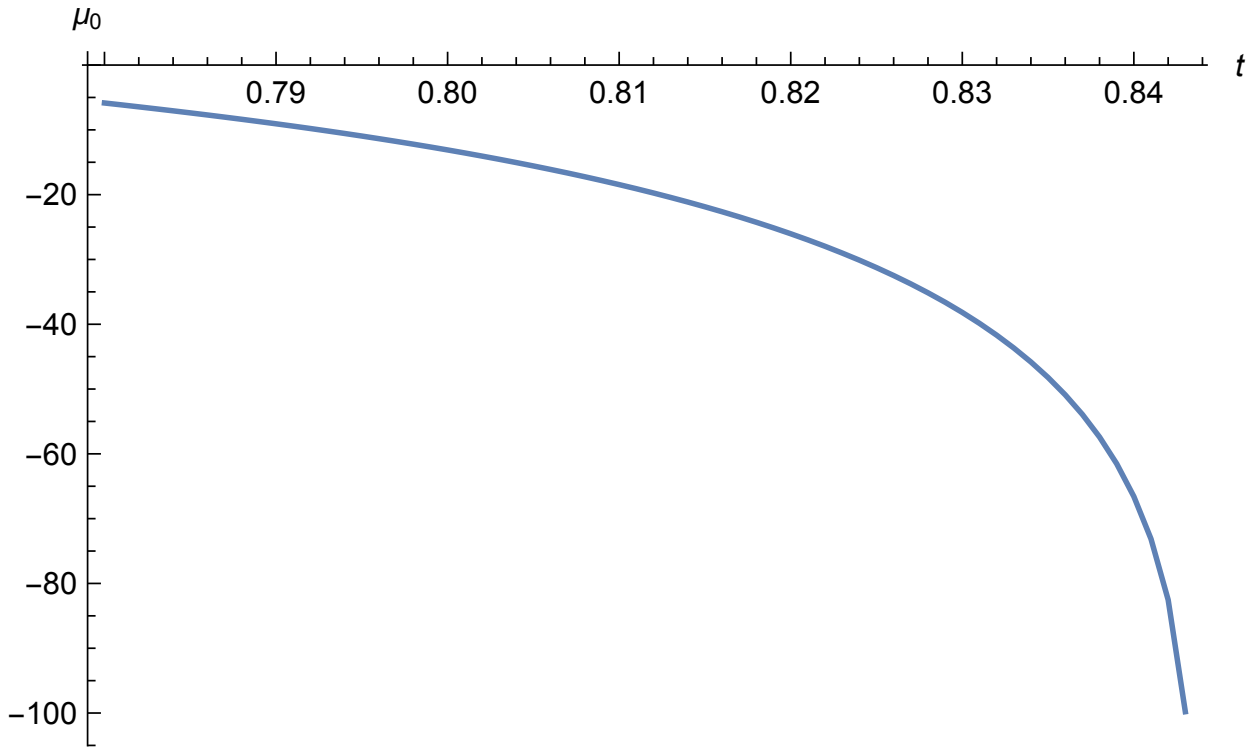}
    \caption{Chemical potential $\mu_0$ versus the temperature $t$
    ($t=T/T_c$, $T_c$ is the critical temperature) for nuclei with mass number $A=50$. }
    \label{fig:03}
\end{figure}

We see that as $a_0$ decreases, the value $\mu_0$ also decreases.
This means that the halo region of the chemical elements with the value $\mu_0$
less than $\mu_0$ of krypton-94 (see Table~\ref{tabl:t1}) becomes wider but more specific.
It becomes a fragmented halo (see~\cite{Chulkov}), i.e., a boiling halo.
The phase transition from liquid to gas occurs starting from the critical point,
and simultaneously the nucleus turns into the nuclear matter
(see Fig.~\ref{fig:03}). The fact that the halo becomes fragmented shows
that the halo becomes really infinitely large.
This is confirmed by the data presented in~\cite{Chulkov}.

Thus, as the halo is fragmented, we deal not with the atomic nucleus but with the nuclear matter
which, in the case of boiling halo, is very similar to the Van-der-Waals (see~\cite{Karnauhov}),
but only  ``turned inside out'', because the temperature becomes higher and higher.
Therefore, the Skyrme forces significantly differ from the Van-der-Waals forces,
because the critical temperature becomes high in the first case,
while it is equal to the unity for the Van-der-Waals gas.

The picture of the halo extension corresponds to the transition from the chemical potential values implying
a wide halo (see Table~\ref{tabl:t1}) to the chemical potential values for the boiling halo.
This phase transition is described by the transition from Fig.~\ref{fig:01} to Fig.~\ref{fig:02},
where the isotherms $T=0.84375$, $T=0.9$, $T=0.95$, and $T=1$ for $\rho>\rho_c$
are shown on the P-Z diagram.
This picture corresponds not to the atomic Bose-nucleus with zero spin but, as was shown above,
to the phase transition from liquid to gas and simultaneously to the phase transition from nucleus
to the nuclear matter. This means the transition from the atomic nucleus
to the nucleus fragmentation and simultaneously to the halo fragmentation,
because the nucleus is related to the halo.

\begin{figure}
    \centering
    \includegraphics[width=0.7\linewidth]{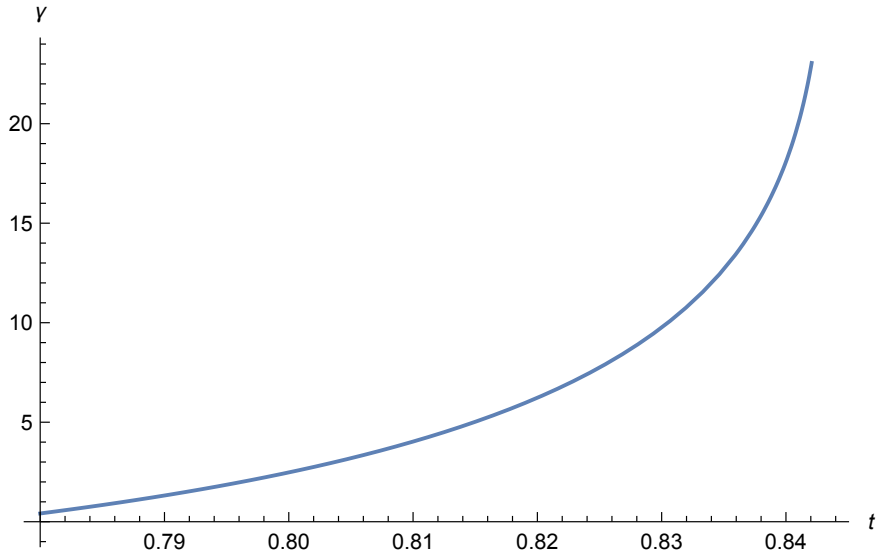}
    \caption{Dependence $\gamma(t)$ for the Van-der-Waals gas.
    The critical temperature is $T_c=16$\,MeV, and the number of nucleons in the nucleus is $A=50$.}
    \label{fig:04}
\end{figure}

\subsection*{Dependence $\gamma(T)$}

We have the expression for the gas pressure
(where $C=\sqrt{m/(2\pi \hbar^2)}$ and $r$ is the nucleus radius):
\begin{equation}\label{ep1}
    P=C^{2(\gamma+1)}T^{2+\gamma} (\operatorname{Li}_{2+\gamma}(a)-\frac{1}{(N+1)^{\gamma+1}}\operatorname{Li}_{2+\gamma}(a^{N+1})).
\end{equation}
On the spinodal ($a=1$), we have
\begin{equation}\label{ep2}
    P_{s}=C^{2(\gamma+1)}T^{2+\gamma} (\operatorname{Li}_{2+\gamma}(1)-\frac{1}{(N_s+1)^{\gamma+1}}\operatorname{Li}_{2+\gamma}(1)).
\end{equation}
Moreover, we know the dependence $N_s(\gamma,T)$ by the formula
\begin{equation}\label{np-1}
    N_s= C^{2(\gamma+1)}{T^{\gamma+1}}r^{2(\gamma+1)} (\operatorname{Li}_{1+\gamma}(1)
    -\frac{1}{(N_s+1)^{\gamma}}\operatorname{Li}_{1+\gamma}(1)),
\end{equation}
Let $p=P/P_c$ and $t=T/T_c$.
We assume that the critical temperature is equal to $T_c=16$\,MeV
(according  to~\cite{Karnauhov}, it lies between 15 and 20\,MeV). Then, on the spinodal, we have
\begin{equation}\label{ep2a}
    p_{s}=t^{2+\gamma} \frac{1-(N_s(\gamma,T)+1)^{-\gamma-1}}{1-{(N_c+1)^{-\gamma-1}}}.
\end{equation}

Equating the negative pressure on the spinodal for the Van-der-Waals gas to the pressure~$p_s$,
$$
p_{sVdW}(t)=-p_s(\gamma_,t),
$$
we obtain the dependence $\gamma(t)$ (see Fig.~\ref{fig:04}).

The case $D=3$ was considered above. One can obtain the function $a_0$ of the temperature $T$
under the assumption that $\gamma$ depends on $T$.
The last dependence can be obtained, for example, from experimental data in the domain of a spinodal.

\subsection*{Dependence $Z(a_0)$}

The formulas of dependence of the pressure $P$ and the number of particles $N$
on the activity for the ideal gas in the case of Gentile statistics are known
($C={m/(2\pi \hbar^2)}$):
\begin{equation}\label{P}
    P=C^{\gamma+1}T^{2+\gamma} (\operatorname{Li}_{2+\gamma}(a)-\frac{1}{(K+1)^{\gamma+1}}\operatorname{Li}_{2+\gamma}(a^{K+1})),
\end{equation}
\begin{equation}\label{N}
    N= {V} C^{\gamma+1}T^{1+\gamma} (\operatorname{Li}_{1+\gamma}(a)-\frac{1}{(K+1)^{\gamma}}\operatorname{Li}_{1+\gamma}(a^{K+1})).
\end{equation}

Let $K=N$. We expand the right-hand side \eqref{N} in the Taylor series in a neighborhood of the point $N=0$
with respect to small $N$ (the parameters $a$, $\gamma$, $C$, and $V$ are fixed):
\begin{equation}     \label{N1}
    \begin{split}
        &N=C^{\gamma+1}V T^{\gamma+1}N (\gamma  \operatorname{Li}_{\gamma +1}(a)
        -\log (a) \operatorname{Li}_{\gamma }(a))+O\left(N^2\right).
    \end{split}
\end{equation}

In the limit as $N\to0$, formula \eqref{N1} implies the equation for $a_0$:
\begin{equation}
    \label{N=0a}
    \gamma  \operatorname{Li}_{\gamma+1}(a_0)-  \log (a_0) \operatorname{Li}_\gamma(a_0)=V^{-1}(C T)^{-\gamma-1}.
\end{equation}

It is well known that $\operatorname{Li}_{\gamma}(a) =a+o(a)$, which leads to the equation
\begin{equation}
\label{a0}
 (\gamma- \log (a_0))a_0+o(a_0\log a_0)-V^{-1}(C T)^{-\gamma-1}=0.
\end{equation}

In turn, the expansion of the pressure~\eqref{P} in the Taylor series in $N$
in a neighborhood of the point $N=0$ has the form
\begin{equation}
\label{E1a}
P=C^{\gamma+1} T^{2+\gamma} N (\gamma  \operatorname{Li}_{\gamma +2}(a)+\operatorname{Li}_{\gamma +2}(a)
-\log (a) \operatorname{Li}_{\gamma +1}(a))+O(N^2).
\end{equation}

Formula~\eqref{E1a} implies the expression for the compressibility factor equal to $Z=\frac{PV}{N T}$:
\begin{equation}
\label{z1}
Z=C^{\gamma+1} V T^{1+\gamma}(\gamma  \operatorname{Li}_{\gamma +2}(a)+\operatorname{Li}_{\gamma +2}(a)
-\log (a) \operatorname{Li}_{\gamma +1}(a))+O(N).
\end{equation}

For $a=a_0$, $N=0$, and hence, \eqref{z1} implies the strict equality
\begin{equation}
\label{z1a0}
Z=C^{\gamma+1} V T^{1+\gamma}(\gamma  \operatorname{Li}_{\gamma +2}(a_0)
+\operatorname{Li}_{\gamma +2}(a_0)-\log (a_0) \operatorname{Li}_{\gamma +1}(a_0)).
\end{equation}
As $a_0\to0$,
\begin{equation}
\label{z11}
Z=C^{\gamma+1} V T^{1+\gamma}((\gamma-1) \ln{a_0} +a_0)+o(a_0\log a_0).
\end{equation}

Formulas \eqref{a0} and \eqref{z11} imply
\begin{equation}
\label{z2}
Z_0=Z(a_0)=C^{\gamma+1} V T^{1+\gamma}(V^{-1}(C T)^{-\gamma-1}+a_0+o(a_0\log a_0))
=1+a_0 C^{\gamma+1} V T^{1+\gamma}+o(a_0\log a_0).
\end{equation}

Since $a_0\gamma=o(a_0\log a_0)$, it follows from \eqref{a0} that
\begin{equation}
\label{a01}
-\log (a_0)a_0+o(a_0\log a_0)-V^{-1}(C T)^{-\gamma-1}=0,
\end{equation}
whence we have
\begin{equation}
\label{a02}
a_0 C^{\gamma+1} V T^{1+\gamma}=\frac{1}{-\log (a_0)}+o(a_0).
\end{equation}

With formula \eqref{a02} taken into account, we see that the compressibility factor
at the point $a_0$ for high temperatures becomes
\begin{equation}
\label{dz}
Z_0=1-\frac{1}{\log {a_0}}+o(a_0)>0.
\end{equation}

Table~\ref{tabl:t2} presents the half-life or the nucleus abundance ($T_{1/2}$),
the neutron separation energy taken from the CDFE data base
($B_{nExp}$ in MeV, which is equal to the nucleus temperature $T$),
the value of $a_0$ calculated by formula \eqref{N=0a},
the value of the chemical potential $\mu_0=T \log{a_0}$ in MeV,
the compressibility factor $Z_0=Z(a_0)$ (by formula \eqref{z1a0}),
and in the last column, the same value but calculated by asymptotic formula~\eqref{dz} ($Z_{0as}$).
\begin{table}[]
\caption{}
    \label{tabl:t2}
    \begin{tabular}{lllllll}
        nucleus  &$T_{1/2}$               &$B_{nExp}$&$a_0$         &$\mu_0$   &$Z_0$    &$Z_{0as}$\\
        26-Fe-50 & 155 ms                 & 17.972   & 0.0000411272 & -181.496 & 1.09433 & 1.09902 \\
        25-Mn-50 & 283.29 ms              & 13.083   & 0.0000696811 & -125.225 & 1.09926 & 1.10448 \\
        24-Cr-50 & 4,345\%                & 13.001   & 0.0000704146 & -124.304 & 1.09936 & 1.10459 \\
        22-Ti-50 & 5.18\%                 & 10.940   & 0.0000939083 & -101.449 & 1.10228 & 1.10784 \\
        20-Ca-50 & 13.9 s                 & 6.354    & 0.000233997  & -53.1207 & 1.11277 & 1.11961 \\
        21-Sc-50 & 102.5 s                & 6.057    & 0.000253733  & -50.1473 & 1.1138  & 1.12078 \\
        18-Ar-50 & \textgreater{}= 170 ns & 4.472    & 0.000424852  & -34.7196 & 1.12084 & 1.1288
    \end{tabular}
\end{table}

\begin{figure}
    \centering
    \includegraphics[width=0.95\linewidth]{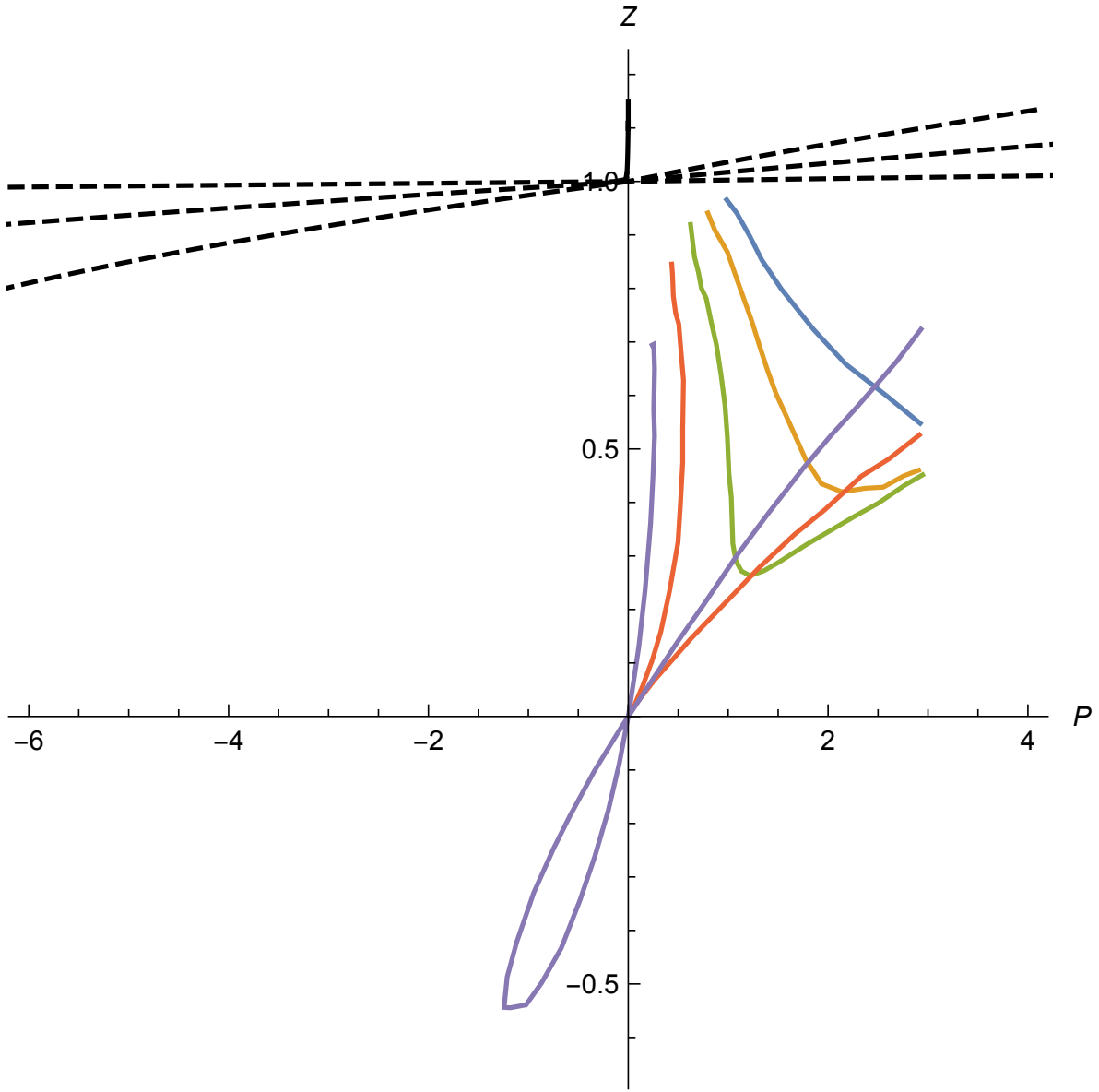}
    \caption{Hougen--Watson diagram. The solid lines are constructed by the P-V diagram
    for the nuclear matter (for the Skyrme model) in accordance with Fig.4 in~\cite{Karnauhov}.
    The dashed lines are the isotherms constructed by formulas \eqref{P} and \eqref{N}
    for nuclei 18-Ar-50, 21-Sc-50, and 25-Mn-50.}
\label{fig:05}
\end{figure}

Let us consider  the  case $K=N$.

Critical parameters will be denoted  by a lower index $c$.
Thus,  for $\mu=0$ we obtain the relations for  $P_c$  and $N_c$:
\begin{equation}\label{Pb}
	P_c=C^{\gamma+1}T_c^{2+\gamma} (\operatorname{Li}_{2+\gamma}(1)-\frac{1}{(N_c+1)^{\gamma+1}}\operatorname{Li}_{2+\gamma}(1)),
\end{equation}
\begin{equation}\label{Nb}
	N_c= \left(\frac{T_c}{\varepsilon}\right)^{1+\gamma} (\operatorname{Li}_{1+\gamma}(1)-\frac{1}{(N_c+1)^{\gamma}}\operatorname{Li}_{1+\gamma}(1)),
\end{equation}
where  $\varepsilon=(C^{\gamma+1} V )^{-\frac{1}{1+\gamma}}$.

We  denote  $x={T_c}/{\varepsilon}$.
From Eq.~\eqref{Nb} one can find $N_c(x,\gamma)$.

The compressibility  factor  has  the  form:
\begin{equation}\label{Z}
	Z_c=\frac{P_c V_c}{N_c T_c}=
\frac{\operatorname{Li}_{2+\gamma}(1)-\frac{1}{(N_c+1)^{1+\gamma}}\operatorname{Li}_{2+\gamma}(1)}
{\operatorname{Li}_{1+\gamma}(1)-\frac{1}{(N_c+1)^{\gamma}}\operatorname{Li}_{1+\gamma}(1)}=
\frac{\zeta(2+\gamma)}{\zeta(1+\gamma)}\frac{1-\frac{1}{(N_c+1)^{1+\gamma}}}{1-\frac{1}{(N_c+1)^{\gamma}}},
\end{equation}

Now we can obtain the  critical value of   the compressibility  factor for the Van-der-Waals  gas:
$Z_{cVdW}=0.375$. Hence  the  value $\gamma_{c}(x)$  can be found  from
\begin{equation}\label{zgx}
\frac{\zeta(2+\gamma)}{\zeta(1+\gamma)}\frac{1-\frac{1}{(N_c(x,\gamma)+1)^{1+\gamma}}}{1-\frac{1}{(N_c(x,\gamma)+1)^{\gamma}}}=Z_{cVdW}.
\end{equation}

\begin{figure}
	\centering
	\includegraphics[width=0.8\linewidth]{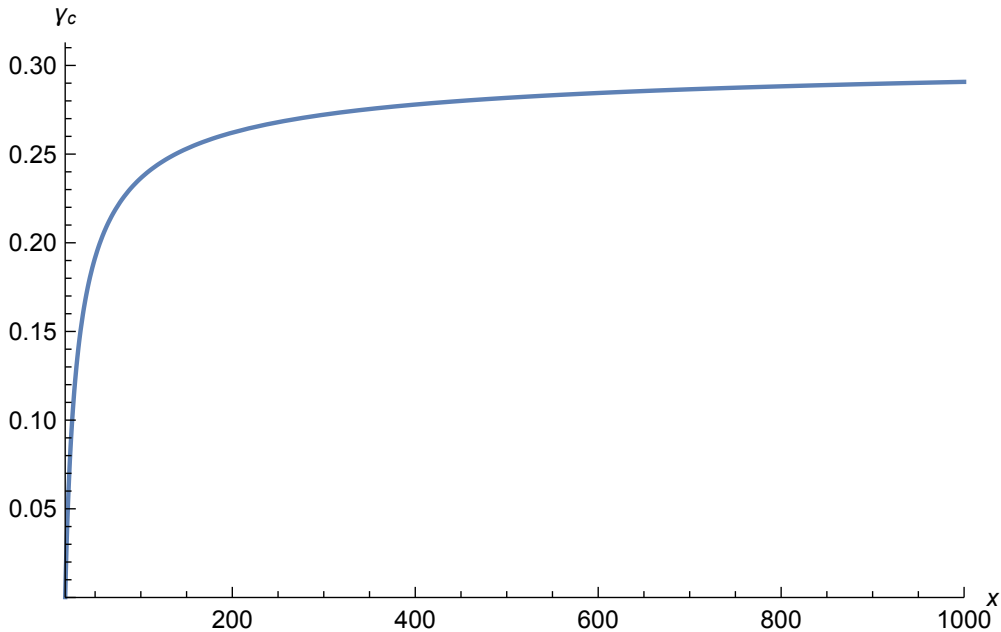}
	\caption{Dependence  of $\gamma_c$ on  $x=\frac{T_c}{\varepsilon}$  on the  critical isothem. It was
found  as a solution of Eq.~\eqref{zgx}.  At infinity the  curve  tends to  a  straight line $\gamma_c=0.312$. This corresponds to the ordinary Bose  gas.}
	\label{fig:06}
\end{figure}

\begin{figure}
	\centering
	\includegraphics[width=0.8\linewidth]{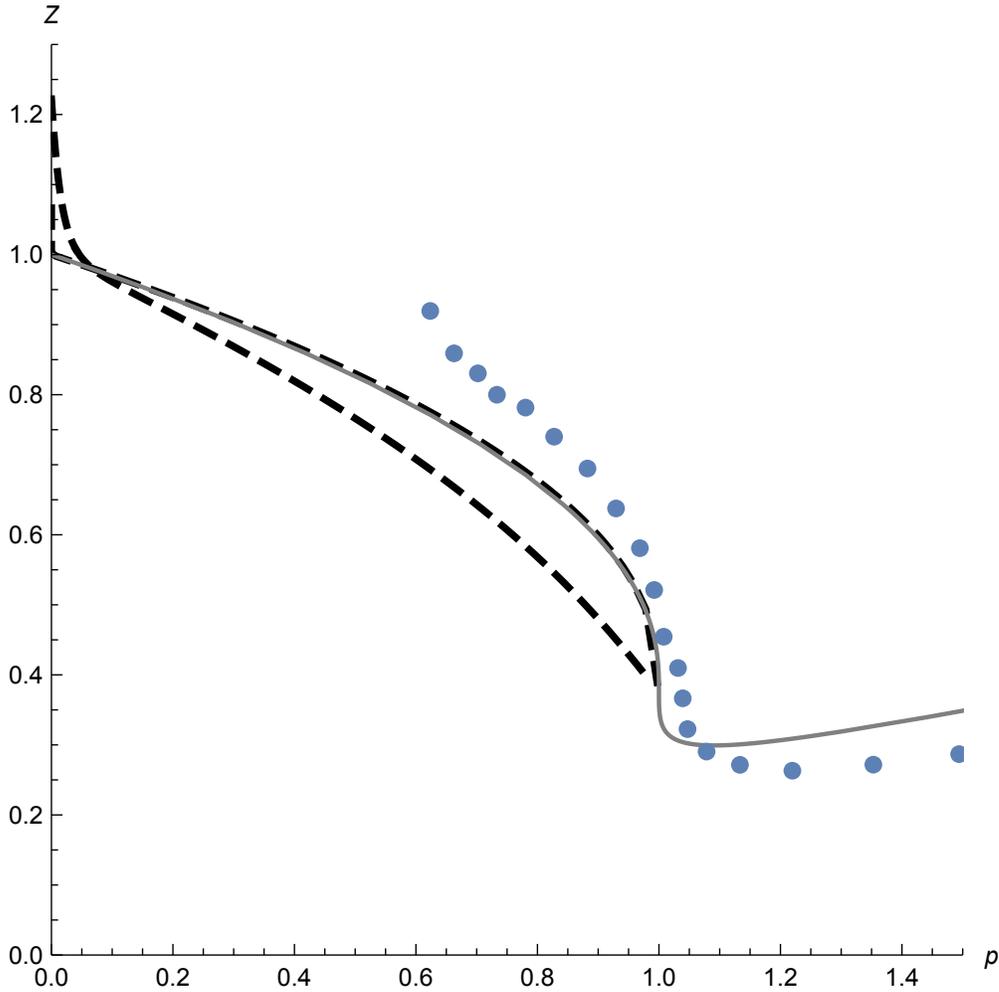}
	\caption{Hougen--Watson diagrams for the  critical isotherm.  $Z=PV/NT$,   $p=P/P_c$.
The points  correspond  to isotherm  $t=T/T_c=1$ on  the P-V diagram
    for the nuclear matter (for the Skyrme model) shown on Fig.4 in~\cite{Karnauhov}.
  The solid  line is the critical isotherm  by the Van-der-Waals model.
  The dashed lines are the isotherms constructed by formulas~\eqref{Pb} and \eqref{Nb} for  $x=17.3$
  (this line reaches  the point $P=0,\, Z=1.22$)
and  for  $x=10000$ (this line reaches  the point $P=0,\,  Z=1.066$). }
\label{fig:07}
\end{figure}

Dependence  of $\gamma_c$ on  $x=\frac{T_c}{\varepsilon}$  on the  critical isothem  is represented in Fig.~\eqref{fig:06}.
In  Fig.~\ref{fig:07} the  Hougen--Watson diagram for the  critical isotherm  is  shown.

\section*{Conclusion}

The separation of one neutron corresponds to the transition of the nucleus of a Bose gas molecule
into the nucleus of a Fermi gas molecule, and conversely.

The passage from particles of a Bose gas to those of a Fermi gas occurs in its roughest form when the activity $a$ changes sign. In
this passage,  the activity $a$ becomes equal to zero, while the chemical potential becomes equal to minus infinity. In this process
one nucleon leaves the nucleus.

We considered the behavior of the Bose--Einstein distribution in some neighborhood of the point $a=0$ and showed that the separation
of a neutron from the nucleus occurs at the point $a=a_0$, which is not zero. Then, using the analog of Gentile statistics for $K=0$,
we calculated the value of the nonstandard specific energy that was needed to separate the nucleon from a particle of the Bose gas.
Although Gentile statistics was previously used for a number of particles greater than 1, the application of nonstandard analysis
(Leibnitz differential or monad)  and  infinitesimal quantities allowed the author to generalize the relations of Gentile statistics
to the case of a small number of Bose particles for $N=K=0$.

The notion of wave packet means that the particle is not point-like, it is diffuse. This process depends on the de Broglie wavelength
of the wave packets.

The main result of the present paper consists in the determination of the value of a parameter $a_0$ that allows us to construct an
antipode of sorts of the Hougen--Watson P-Z diagram for nuclear matter. We have shown that, knowing the values of $a_0$, we can
construct all the isotherms on the P-Z diagram.

The Bohr model of the nucleus of an atom is similar to the model of liquids. In the author's
papers~\cite{RJMP_22-3},~\cite{RJMP_22-1}, it was shown that a liquid complying with the Van-der-Waals model may be approximated by
the Bose--Einstein distribution for negative pressures and by the Fermi--Dirac distribution for positive ones. For a pressure $P< 0$
the liquid dilates. The dilation of liquids at negative pressures was already noticed by Huygens. It turns out that in the region of
positive pressure the Fermi--Dirac distribution describes the picture of a Van-der-Waals gas sufficiently well. This connection
indirectly confirms the analogy between nuclear matter and liquids.

\end{document}